\documentclass[english,prl,aps,floatfix,superscriptaddress,twocolumn,longbibliography]{revtex4-1}
\usepackage[T1]{fontenc}
\usepackage[latin9]{inputenc}
\usepackage{color}
\usepackage{float}
\usepackage{bm}
\usepackage{amsmath}
\usepackage{amssymb}
\usepackage{graphicx}
\usepackage{wasysym}
\usepackage{cancel}

\makeatletter

\usepackage{amssymb}
\usepackage{amsmath}
\usepackage{graphicx}
\usepackage[colorlinks=true,linkcolor=blue,anchorcolor=red,citecolor=blue,urlcolor=blue]{hyperref}

\makeatother

\usepackage{babel}
\begin{document}

\title{Chirality Josephson current due to a novel quantum anomaly in inversion-asymmetric Weyl semimetals}

\author{Song-Bo Zhang}

\affiliation{Institute for Theoretical Physics and Astrophysics, University of
Würzburg, D-97074 Würzburg, Germany}

\author{Johanna Erdmenger}

\affiliation{Institute for Theoretical Physics and Astrophysics, University of
Würzburg, D-97074 Würzburg, Germany}

\author{Björn Trauzettel}

\affiliation{Institute for Theoretical Physics and Astrophysics, University of
Würzburg, D-97074 Würzburg, Germany}

\date{\today}
\begin{abstract}
We study Josephson junctions based on inversion-asymmetric but time-reversal symmetric Weyl semimetals under the influence of Zeeman fields. We find that, due to distinct spin textures, the Weyl nodes of opposite chirality respond differently to an external magnetic field. Remarkably, a Zeeman field perpendicular to the junction direction results in a phase shift of opposite sign in the current-phase relations of opposite chirality. This leads to a finite chirality Josephson current (CJC) even in the absence of a phase difference across the junction. This feature could allow for applications in chiralitytronics. In the long junction and zero temperature limit, the CJC embodies a novel quantum anomaly of Goldstone bosons at $\pi$ phase difference which is associated with a $\mathbb{Z}_2$ symmetry at low energies. It can be detected experimentally via an anomalous Fraunhofer pattern.
\end{abstract}
\maketitle
\textit{Introduction.\textendash }Weyl semimetals (WSMs) have recently attracted intensive interest thanks to their realization in a number of materials \cite{Hosur13Physique,Turner2013arXiv,Burkov11prl,Halasz12PRB,Hirayama14PRL,Weng15prx,Huang15nc,Rauch15PRL,Xu15sci-TaAs,Yang15Nphys,Lv15prx,Lv15nphys,Xu15np-NbAs,XuN16NC-TaP,Ruan16Ncomms,Ruan16PRL}
and to their novel physics associated with Weyl nodes \cite{Nielsen83plb,Xu11prl,Yang11prb,Zyuzin12prb,Hosur12prl,Son13prb,Vazifeh13prl,Burkov14prl,Potter14nc,Burkov14prl-chiral,Parameswaran14prx,Lu15Weyl-shortrange,Landsteiner14PRB,ZhouJH15prb,ZhangSB16NJP,ZhangCL16nc,Li16np}.
The Weyl nodes appear in pairs that carry opposite chirality \cite{Nielsen81plb} in the absence of time-reversal or inversion symmetry. Chirality is thus a defining emergent property of electrons in WSMs. The possibility to probe and manipulate chirality is one of the most intriguing aspects of WSMs. Recently, chirality-dependent physics has been discussed in various contexts \cite{YangSY15PRL,Jiang15PRL,Chan16PRL}.

Josephson junctions provide a complementary method to probe the electronic properties of a particular material. They are the basic building blocks for superconducting electronics with applications ranging from electronic magnetometers to quantum computation \cite{Baselmans99nature,Golubov04RMP,Devoret13science,Gambetta17Nature}.
Hence, it is of fundamental interest to study Josephson junctions based on WSMs. To date, most experimentally relevant
WSMs preserve time-reversal symmetry but break inversion symmetry
\cite{Rauch15PRL,Xu15sci-TaAs,Lv15prx,Lv15nphys,XuN16NC-TaP,Weng15prx,Huang15nc,Hirayama14PRL,Yang15Nphys,Xu15np-NbAs,Ruan16Ncomms,Ruan16PRL}. In these materials, s-wave superconductivity couples electrons of the same chirality \cite{Meng12PRB,Songbo18PRB}. Thus, chirality remains a well-defined property in those Josephson junctions. Hence, we could think about using chirality as a resource for electronics, just as spin in spintronics. We coin this idea chiralitytronics. Recently, Josephson junctions based on Dirac semimetals have been fabricated \cite{WYu18PRL,CZLi18PRB,CL17arXiv}.  By similar methods, it is feasible to also investigate Josephson junctions based on WSMs. Previous theoretical work  \cite{Kim16PRB,Khanna16PRB,Madsen17PRB,Bovenzi17PRB,Chen17PRB} instead focused on either the inversion-symmetric case or the surface states where chirality is no longer a good quantum number \cite{Meng12PRB,Cho12PRB}.

In this Letter, we focus on Josephson junctions to study the interplay of Zeeman fields, s-wave superconductivity, and chirality in inversion-asymmetric WSMs. We find that the Weyl nodes of opposite chirality display distinct spin textures. Thus, they respond differently to an external magnetic field. As a result, a Zeeman field perpendicular to the junction direction shifts the phase in the current-phase relations (CPRs) of opposite chirality differently. For each chirality, a Josephson $\phi_{0}$-junction with a phase shift of opposite sign for opposite chirality is realized. The phase shift is controllable by the junction length and external Zeeman field. The relations between the CPRs can be understood by the underlying low-energy symmetries of the system. Remarkably, this mechanism gives rise to the phenomenon of a finite CJC $J_s^{\text{chi}}\equiv J_s^{+}-J_s^{-}$ with $J_s^{\pm}$ the Josephson current for each chirality.  In the long junction and zero temperature limit, we show that this phenomenon expresses a novel quantum anomaly of Goldstone bosons (Cooper pairs) at $\pi$ phase difference, since when sending the Zeeman field to zero, the CJC retains a sign singularity, breaking the $\mathbb{Z}_{2}$ symmetry between the two decoupled chirality sectors at the quantum level. This mechanism also manifests itself as an anomalous Fraunhofer pattern.

\begin{figure}[H]
\centering

\includegraphics[width=8cm]{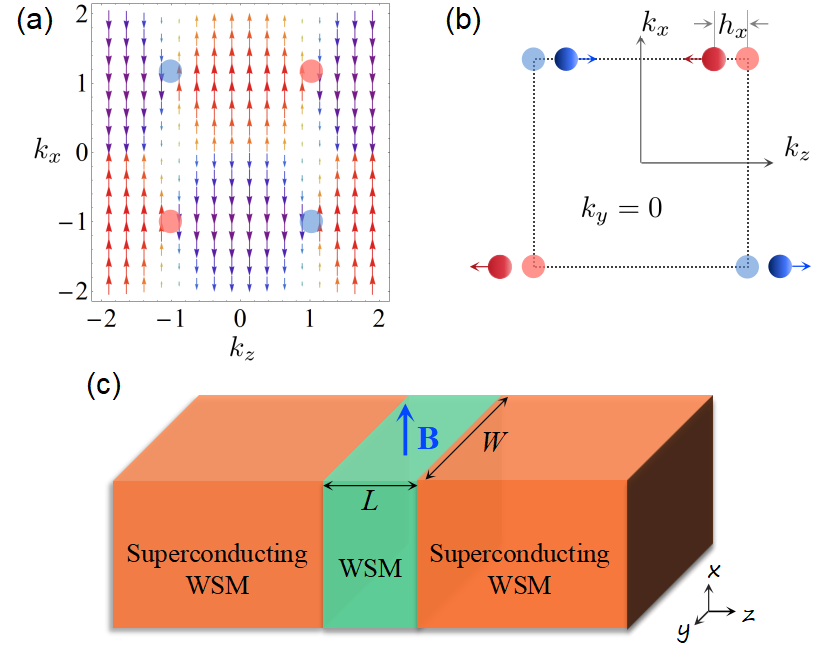}

\caption{(a) Spin texture of the lower conduction band of the model (\ref{eq:StartingModel}) at $k_{y}=0$ with $\kappa_{0}=\sqrt{2}$ and $\beta=1$;  (b) Position shifts of Weyl nodes in momentum space due to the Zeeman field $h_{x}$ in $x$ direction. The red (blue) dots denote the Weyl nodes of positive (negative) chirality; (c) Sketch of the Weyl
Josephson junction. }

\label{Fig:Wnode-shift}
\end{figure}

\textit{Model\ and\ setup.\textendash }We consider inversion-asymmetric WSMs described by the Hamiltonian\ \cite{Songbo18PRB,Kourtis16PRB}
$\mathcal{H}_{\text{w}}=\sum_{{\bf {\bf k}}}\psi_{{\bf {\bf k}}}^{\dagger}H({\bf k})\psi_{{\bf {\bf k}}}$
with
\begin{align}
H({\bf k}) & =k_{x}\sigma_{x}s_{z}+k_{y}\sigma_{y}s_{0}+(\kappa_{0}^{2}-|{\bf k}|^{2})\sigma_{z}s_{0}\nonumber \\
 & \ \ \ +\beta \sigma_{y}s_{y}-\alpha k_{y}\sigma_{x}s_{y},\label{eq:StartingModel}
\end{align}
where $\psi_{{\bf k}}^{\dagger}=(c_{A,\uparrow,{\bf k}}^{\dagger},c_{A,\downarrow,{\bf k}}^{\dagger},c_{B,\uparrow,{\bf k}}^{\dagger},c_{B,\downarrow,{\bf k}}^{\dagger})$, $c_{\sigma,s,{\bf k}}^{\dagger}$ are creation operators with orbital indices $\sigma=A,B$, spin indices $s=\uparrow,\downarrow$, and wave vector ${\bf k}$; $\sigma_{i}$ ($i=x,y,z,0$) are Pauli and $2\times2$ unit matrices for orbital space, and $s_{i}$ for spin space. $\kappa_{0},$ $\alpha$ and $\beta$ are real model parameters. Suppose $0$$<$$\beta$$<$$\kappa_{0}$, there are four Weyl nodes at ${\bf Q}_{1,2}=\pm\left(\beta,0,k_{0}\right)$ and ${\bf Q}_{3,4}=\pm\left(\beta,0,-k_{0}\right)$, respectively, where $k_{0}$$=$$(\kappa_{0}^{2}-\beta^{2})^{1/2}$. At low energies, the model (\ref{eq:StartingModel}) can be approximated as a sum of four Weyl Hamiltonians, $\mathcal{H}_{\text{w}}=\sum_{\gamma=1}^{4}\sum_{{\bf k}}'\Psi_{\gamma,{\bf k}}^{\dagger}H_{\gamma}({\bf k})\Psi_{\gamma,{\bf k}}$ with
\begin{align}
 & \begin{array}{c}
H_{1(2)}({\bf k})=\left(k_{x}\mp\beta\right)s_{x}+k_{y}s_{y}+\left(k_{z}\mp k_{0}\right)s_{z},\\
H_{3(4)}({\bf k})=\left(k_{x}\mp\beta\right)s_{x}+k_{y}s_{y}-\left(k_{z}\pm k_{0}\right)s_{z},
\end{array}\label{eq:subHamiltonians}
\end{align}
where $k_{y}$ has been rescaled by $1/\alpha$ and $k_{z}$ by $1/(2k_{0})$ \ \cite{Songbo18PRB}. $\gamma(=1,2,3,4)$ labels the Weyl node at ${\bf Q}_{\gamma}$. Accordingly, Weyl nodes $1$ and $2$ carry positive chirality, while Weyl nodes 3 and 4 carry negative chirality. $\sum_{{\bf k}}'$ means that ${\bf k}$ is confined in the sum to the vicinity of Weyl nodes. The spinors
$\Psi_{\gamma,{\bf k}}^{\dagger}\equiv(\psi_{\gamma,\uparrow,{\bf k}}^{\dagger},\psi_{\gamma,\downarrow,{\bf k}}^{\dagger})$ of Weyl nodes are given by $\Psi_{1,{\bf k}}^{\dagger}=\Psi_{3,{\bf k}}^{\dagger}=(c_{\uparrow,{\bf k}}^{(B)\dagger},c_{\downarrow,{\bf k}}^{(A)\dagger})$
and $\Psi_{2,{\bf k}}^{\dagger}=\Psi_{4,{\bf k}}^{\dagger}=(c_{\uparrow,{\bf k}}^{(A)\dagger},c_{\downarrow,{\bf k}}^{(B)\dagger})$ with $c_{\uparrow(\downarrow),{\bf k}}^{(\sigma)}=(c_{\sigma,\uparrow,{\bf k}}\pm c_{\sigma,\downarrow,{\bf k}})/\sqrt{2}$. While the indices $s=\uparrow,\downarrow$ in the operators $c_{\sigma,s,{\bf k}}$ denote spin-up and spin-down in $z$ direction, respectively, $s'=\uparrow,\downarrow$ in the new basis $\psi_{\gamma,s',{\bf k}}=c_{s',{\bf k}}^{(\sigma)}$ denote spin-up and spin-down in $x$ direction, respectively. We can readily observe that Weyl nodes of opposite chirality display distinct spin textures or spin-momentum locking as shown in Fig.\ \ref{Fig:Wnode-shift}(a).

The distinct spin textures near Weyl nodes of opposite chirality imply different responses of Weyl nodes to Zeeman fields. A Zeeman field ${\bf h}=g\mu_{B}{\bf B}/2\equiv(h_{x},h_{y},h_{z})$ couples to the electron spin via $H_{\text{Z}}=\sigma_{0}{\bf h}\cdot\bm{s}$ in our model\ (\ref{eq:StartingModel}), where $g$ is the g-factor, $\mu_{B}$ the Bohr magneton and ${\bf B}$ the magnetic field. The $y$- and $z$-components $h_{y}$ and $h_{z}$ couple electrons from different Weyl nodes, whereas the $x$-component $h_{x}$ couples electrons not only from different Weyl nodes but also acts within each Weyl node. Therefore, after projecting ${\bf h}$ to low energies, only $h_{x}$ is significant\ \cite{Supplementary,Note2}. It gives rise to a diagonal term
\begin{align}
H_{\gamma}^{\text{Z}} & =h_{x}s_{z}\label{eq:Zeeman-term}
\end{align}
for each Weyl node. Combining Eq.\ (\ref{eq:Zeeman-term}) with Eq.\ (\ref{eq:subHamiltonians}),
it is interesting to see that the positions of the Weyl nodes of positive (negative) chirality are shifted oppositely by $\mp h_{x}$ in $k_{z}$ direction, as depicted in Fig.\ \ref{Fig:Wnode-shift}(b). Thus, $h_{x}$ can be viewed as a constant axial-vector potential $A_{z}^{a}=\hbar h_{x}/e$  in $z$ direction which acts with opposite sign on Weyl nodes of opposite chirality.

Next, we introduce s-wave superconducting pairing with both intra- and interorbital components to this problem. As shown in Ref.\ \cite{Songbo18PRB}, at low energies only the intraorbital pairing is important and it couples Weyl nodes of the same chirality. Thus, the full system can be considered at low energies as two effectively decoupled sectors with opposite chirality, respectively.

Finally, we consider a Josephson junction formed by sandwiching a WSM between two s-wave superconducting WSMs, as sketched in Fig.\ \ref{Fig:Wnode-shift}(c). The positive chirality sector can then be described by the Bogoliubov-de Gennes (BdG) Hamiltonian \cite{Note3}
\begin{align}
h_{\text{BdG}}^{+}= & \nu_{z}\tau_{0} [-i{\bf \partial}_{{\bf r}}\cdot\bm{s}-\mu({\bf r})s_{0}]+h({\bf r})\nu_{0}\tau_{0}s_{z}\nonumber \\
 & +\Delta_{s}({\bf r})e^{i\text{sgn}(z)\phi\nu_{z}/2}\nu_{x}\tau_{x}s_{0}\label{eq:BdG-Hamiltonian-text}
\end{align}
in the Nambu basis $(\psi_{1,\uparrow},\psi_{1,\downarrow},\psi_{2,\downarrow}^{\dagger},-\psi_{2,\uparrow}^{\dagger},\psi_{2,\uparrow},\psi_{2,\downarrow},\psi_{1,\downarrow}^{\dagger},$ $-\psi_{1,\uparrow}^{\dagger})$, $\Delta_{s}({\bf r})=\Delta\Theta(|z|-L/2)$
is the pairing potential, $\phi$ the phase difference across the junction and $L$ the junction length; $\Theta(z)$ is the Heaviside function and $\text{sgn}(z)$ the sign function. The Zeeman field is applied  to the normal WSM in $x$ direction, $h({\bf r})=h_{x}\Theta(L/2-|z|)$; $\mu({\bf r})=\mu_{S}\Theta(|z|-L/2)+\mu_{N}\Theta(L/2-|z|)$ is the chemical potential. The Pauli and $2\times2$ unit matrices $\tau_{i}$ and $\nu_{i}$~($i$$=$$x,y,z,0$) act on Weyl-node and particle-hole spaces, respectively. The BdG Hamiltonian $h_{\text{BdG}}^{-}$ for the negative chirality sector can be obtained by replacing $\partial_{z}$ by $-\partial_{z}$ in Eq.\ (\ref{eq:BdG-Hamiltonian-text})

\textit{Josephson current of one chirality sector.\textendash }In order to determine the Josephson current, we adapt the method of Refs.\ \cite{Brouwer97SSC,Dolcini07PRB,Beenakker13PRL} which takes into account both contributions from Andreev bound states and the continuum spectrum. Throughout the Letter, we normalize the current densities by the corresponding normal-state resistance $R_{n}$ of the junction at zero temperature\ \cite{Supplementary}.

\begin{figure}[h]
\centering

\includegraphics[width=8cm]{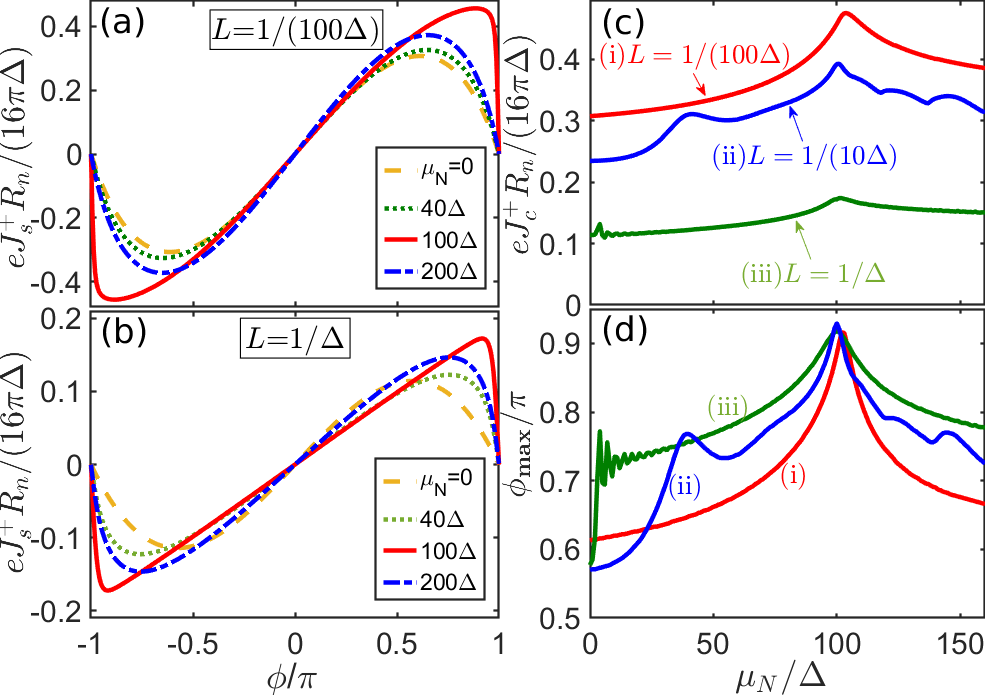}

\caption{Current-phase relations for positive chirality in the absence of Zeeman
fields for $L=1/(100\Delta)$ (a) and $1/\Delta$ (b) with various
choices of $\mu_{N}$; (c) Critical current density $J_{c}^{+}$ as
a function of $\mu_{N}$ for $L=1/(100\Delta)$ (i, red), $1/(10\Delta)$
(ii, blue), and $1/\Delta$ (iii, green), respectively; (d) Location
$\phi_{\text{max}}$ of $J_{c}^{+}$,
a measurement of skewness, as a function of $\mu_{N}$ for (i), (ii),
and (iii), respectively. $\mu_{S}=100\Delta$ and $k_{B}T=0.01\Delta$
in all plots. }

\label{Fig:JosepshonCurrent-phi}
\end{figure}

First, let us analyze the case without Zeeman fields. There, we find that the Josephson currents of the two chiralities are identical, reflecting a $\mathbb{Z}_{2}$ symmetry of the system at low energies which we further discuss below. The results for positive chirality at low temperature are shown in Fig.\ \ref{Fig:JosepshonCurrent-phi}. The Josephson current is $2\pi$-periodic in the phase difference $\phi$. It vanishes at $\phi=0$ and $\pm\pi$ as required by time-reversal symmetry. The critical current density $J_{c}^{+}\equiv\max[J_{s}^{+}(\phi)]$ decays on increasing $L$. At $\mu_{N}\approx\mu_{S}$, $J_{c}^{+}R_{n}$ is maximized {[}Fig.\ \ref{Fig:JosepshonCurrent-phi}(c){]}, indicating the enhancement of Andreev reflection at the N-S interfaces \cite{Songbo18PRB}. Interestingly, we predict a clear forward skewness in the CPRs at low temperatures $k_{B}T\ll\text{min}\{\Delta,1/L\}$. This skewness is related to the helical nature of Andreev bound states \cite{Sochnikov15PRL,Tkachov15PRB} due to strong spin-momentum locking in WSMs. At $\mu_{N}\approx\mu_{S}$, a more pronounced skewness can be observed {[}Fig.\ \ref{Fig:JosepshonCurrent-phi}(d){]}.

\begin{figure}[h]
\centering

\includegraphics[width=8cm]{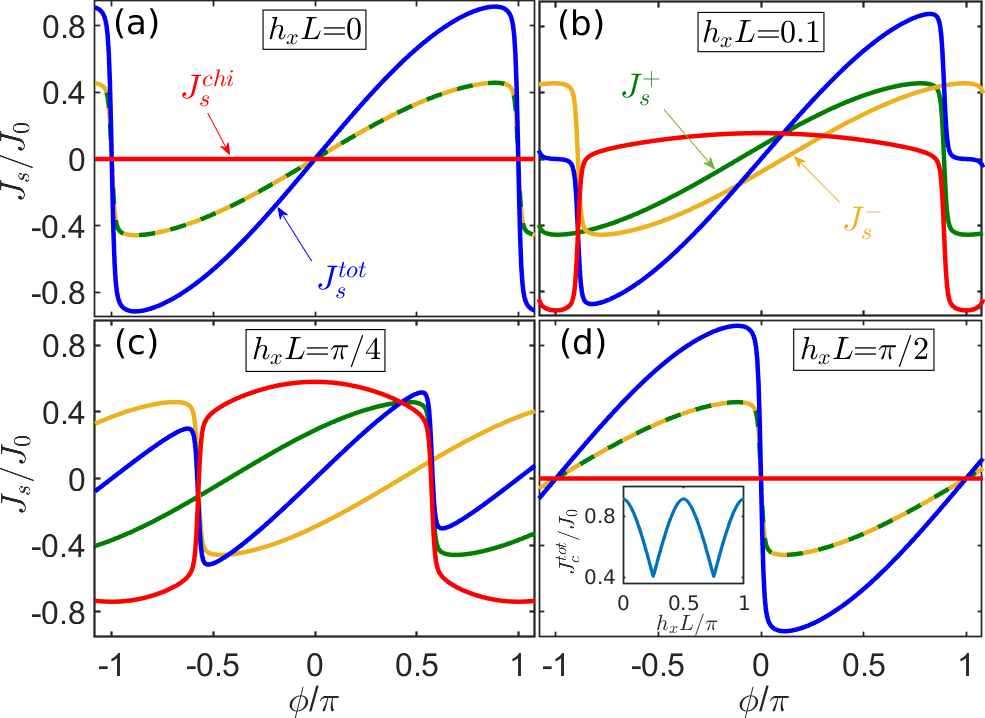}

\caption{Current-phase relations for $h_{x}L=0,$ 0.1, $\pi/4$, and $\pi/2$, respectively. The green, yellow, blue and red curves represent $J_{s}^{+}$, $J_{s}^{-}$, $J_{s}^{\text{tot}}$ and $J_{s}^{\text{chi}}$, respectively. The inset in panel (d) illustrates the maximum of $J_s^{\text{tot}}$ as a function of $h_xL$.
$J_0=16\pi\Delta/(eR_n)$, $L=1/(100\Delta)$, $\mu_{N}=\mu_{S}=100\Delta$ and $k_{B}T=0.01\Delta$
in all plots.}

\label{Fig:CPRl-chiral}
\end{figure}

The most exciting physics arises in the presence of the Zeeman field $h_{x}$. We find that $h_{x}$ shifts the phase difference $\phi$ by $\phi_{0}=2h_{x}L$ as it enters as an effective vector potential in $z$ direction. Consequently, $J_{s}^{+}$ becomes zero neither at $\phi=0$ nor $\pm\pi$. An anomalous Josephson current $J_{s}^{+}(\phi=0)\neq0$ can be obtained, due to the presence of the phase shift $\phi_{0}$ (green curves in Fig.\ \ref{Fig:CPRl-chiral}). The system indeed realizes a Josephson $\phi_{0}$-junction \cite{Dolcini15PRB} for positive chirality with the phase shift tunable by the Zeeman field and junction length. Importantly, similar results can be found for negative chirality but with a phase modulation of opposite sign $-\phi_{0}$ (yellow curves in Fig.\ \ref{Fig:CPRl-chiral}). Therefore, the supercurrents of opposite chirality become different, i.e., $J_{s}^{-}(\phi)\neq J_{s}^{+}(\phi)$. This implies the striking phenomenon of a finite CJC. Note that supercurrents of opposite chirality can still be related to each other by $J_{s}^{-}(\phi)=-J_{s}^{+}(-\phi)$, as a result of a magnetic $\mathbb{Z}_{2}$ symmetry between the two chirality sectors\ \cite{Supplementary}.

\textit{Total and chirality Josephson currents.\textendash }With the obtained $J_{s}^{\pm}(\phi)$, we are equipped to evaluate the total and chirality Josephson currents. The results for the total Josephson current $J_{s}^{\text{tot}}(\phi)\equiv J_{s}^{+}(\phi)+J_{s}^{-}(\phi)$ are represented by the blue curves in Fig.\ \ref{Fig:CPRl-chiral}. In the absence of $\phi_{0}$, the CPRs for each chirality are identical. Thus, $J_{s}^{\text{tot}}(\phi)$ takes exactly the same form as $J_{s}^{+}(\phi)$ but twice as large. In contrast, the presence of $\phi_{0}$ shifts the CPRs of opposite chirality oppositely, which, along with the forwardly skewed shape of $J_{s}^{\pm}(\phi)$, leads to several particular features in $J_{s}^{\text{tot}}(\phi)$. First, the contributions of the two chiralities to $J_{s}^{\text{tot}}(\phi)$ is controllable by $\phi_{0}$. At $\phi=\pm\phi_{0}+n\pi,$ $n\in\mathbb{Z}\equiv\{0,\pm1,...\}$, $J_{s}^{\text{tot}}(\phi)$ is of purely positive (negative) chirality. This feature can be exploited in superconducting chiralitytronics, since we are able to transfer a net chirality across the junction via dissipationless transport of chiral Cooper pairs.
Second, with increasing $\phi_{0}$ from 0 to $\pi/2$, the maximum of $J_{s}^{\text{tot}}(\phi)$
decreases monotonically but never goes to zero {[}inset in Fig.\ \ref{Fig:CPRl-chiral}(d){]}. Meanwhile, the slope of $J_{s}^{\text{tot}}(\phi)$ at $\phi=(2n+1)\pi$, $n\in\mathbb{Z}$, changes its sign at a critical $\phi_{0}$. Then, two instead of one peak (dip) appear in a period $\phi\in[-\pi,\pi]$ {[}Fig.\ \ref{Fig:CPRl-chiral}(b){]}.  Third, for $\phi_{0}=(2n+1)\pi/2$, $n\in\mathbb{Z}$, the two peaks (dips) become equal. Hence, $J_{s}^{\text{tot}}(\phi)$ becomes $\pi$ instead of $2\pi$ periodic in $\phi$ and takes a more skewed shape {[}Fig.\ \ref{Fig:CPRl-chiral}(c){]}. Fourth, for $\phi_{0}=n\pi$, $n\in\{\pm1,\pm3,...\}$, $J_{s}^{\text{tot}}(\phi)$ resembles the one of a $\pi$-junction {[}Fig.\ \ref{Fig:CPRl-chiral}(d){]}. This means that tuning $\phi_{0}$ leads to a $0\text{-}\pi$ transition in $J_{s}^{\text{tot}}(\phi)$. Finally, although time-reversal symmetry is broken by $h_{x}$ in the system, $J_{s}^{\text{tot}}(\phi)$ still obeys $J_{s}^{\text{tot}}(\phi)=-J_{s}^{\text{tot}}(-\phi)$. Thus, $J_{s}^{\text{tot}}(\phi)$ always vanishes at $\phi=n\pi$,
$n\in\mathbb{Z}$.

The red curves in Fig.\ \ref{Fig:CPRl-chiral} show the results for the CJC $J_{s}^{\text{chi}}(\phi)\equiv J_{s}^{+}(\phi)-J_{s}^{-}(\phi)$. Without Zeeman fields, i.e, $\phi_{0}=0$, $J_{\text{\ensuremath{s}}}^{\text{chi}}(\phi)$ is always vanishing, as the supercurrents of opposite chirality $J_{s}^{\pm}(\phi)$ are the same {[}Fig.\ \ref{Fig:CPRl-chiral}(a){]}. In contrast, an applied Zeeman field $h_{x}$ in the N region leads to chirality-dependent phase shifts $\pm\phi_{0}$ in the CPRs, making $J_{s}^{\pm}(\phi)$ generally different. Thus, a finite CJC $J_{\text{\ensuremath{s}}}^{\text{chi}}(\phi)\neq0$ is allowed, as shown in Fig.\ \ref{Fig:CPRl-chiral}(b, c). Different from the total Josephson current, $J_{\text{\ensuremath{s}}}^{\text{chi}}(\phi)$
is always $2\pi$-periodic in $\phi$. Moreover, $J_{\text{\ensuremath{s}}}^{\text{chi}}(\phi)$
is an even function of $\phi$, i.e., $J_{\text{\ensuremath{s}}}^{\text{chi}}(\phi)=J_{\text{\ensuremath{s}}}^{\text{chi}}(-\phi)$. This indicates that a finite $J_{\text{\ensuremath{s}}}^{\text{chi}}$ exists even in the absence of a phase difference. Interestingly, the maxima of $|J_{\text{\ensuremath{s}}}^{\text{chi}}(\phi)|$ always occur at $\phi=n\pi$, $n\in\mathbb{Z}$ (where $J_{s}^{\text{tot}}=0$) and are twice as large as $|J_{s}^{\pm}(\phi)|$ there. In the case $\mu_{N}\approx\mu_{S}$, salient dips may occur at $\phi=(2n+1)\pi$, $n\in\mathbb{Z}$ for small $\phi_{0}$, due to the pronouncedly skewed shape of $J_{s}^{\pm}(\phi)$. Additionally, the anomalous CJC $J_{\text{\ensuremath{s}}}^{\text{chi}}(0)$ is an odd and oscillatory function of $\phi_{0}$ with a period of $2\pi$.

\textit{Quantum anomaly of Goldstone bosons.\textendash }In the Weyl Josephson junction without Zeeman fields, the two chirality sectors at low energies are connected by a $\mathbb{Z}_{2}$ exchange symmetry,
\begin{equation}
\mathcal{U}H_{+}({\bf r})\mathcal{U}^{-1}=H_{-}({\bf r}),\ \ \mathcal{U}=i\tau_{y}s_{y}\mathcal{R}_{x},\label{eq:symmetry}
\end{equation}
where the basis is $(c_{\uparrow}^{(B)},c_{\downarrow}^{(A)},c_{\uparrow}^{(A)},c_{\downarrow}^{(B)})$,
$H_{+}({\bf r})=\text{diag}(H_{1}({\bf r}),H_{2}({\bf r}))$ and $H_{-}({\bf r})=\text{diag}(H_{3}({\bf r})$,
$H_{4}({\bf r}))$ the Hamiltonians for Weyl fermions of positive and negative chirality, respectively, $H_{\gamma}({\bf r})$ the Fourier transforms of $H_{\gamma}({\bf k})$ in Eq.\ (\ref{eq:subHamiltonians}), and $\mathcal{R}_{x}$ the reflection operator about the $yz$ plane\ \cite{Supplementary}. The two chirality sectors have identical response to a phase difference $\phi$ across the junction, as superconductivity preserves this $\mathbb{Z}_{2}$ symmetry. This leads to $J_{s}^{+}(\phi)=J_{s}^{-}(\phi)$ for all $\phi$. Therefore, $J_{s}^{\text{chi}}(\phi)$ is vanishing. However, the Zeeman field $h_{x}$ plays the role of an axial-vector potential in the system. It effectively modifies the phase difference, leading to a finite CJC.    This CJC is closely associated with the $\mathbb{Z}_2$ symmetry, different from the anomalous Josephson effects  discussed previously in a variety of 1D systems with broken chiral symmetry \cite{Krive04LTP,Dolcini15PRB}.

We now carefully explain in which sense the presence of a finite $J_s^{\text{chi}}$ corresponds to a quantum anomaly of Goldstone bosons. In the long junction and zero temperature limit, $T=0$ and $ L \gg 1/\Delta$, the contribution to the supercurrent comes from low-energy Andreev bound states with energy $|\epsilon|\ll\Delta$ \cite{Bardeen72PRB,Maslov96PRB}. For each $\bf{k}_\parallel$, the energy levels cross at $\phi \pm \phi_0=\pi$ for the positive and negative chirality sectors, respectively. Summing over all allowed $\bf{k}_\parallel$, the CJC for fixed $\phi=\pi$ shows a discontinuous jump at $h_x=0$ \ \cite{Supplementary}. Namely,
\begin{equation}
\label{distoninuity}
  J_s^{\text{chi}}= \dfrac{2\mu_N^2}{3L}\Big[\dfrac{2h_xL}{\pi}-\text{sgn}(h_x)\Big], \ \  |h_xL|< \pi.
\end{equation}
This result is robust against a non-magnetic perturbations such as an interface barrier or smooth disorder in the N region  \ \cite{Supplementary}. The amplitude of $J_s^{\text{chi}}$ is determined by the density of states of Weyl fermions at the Fermi level in the N region, and it depends quadratically on $\mu_N$. When $h_x \rightarrow 0$, the system is invariant under the $\mathbb{Z}_2$ exchange. However, $J_s^{\text{chi}}$ retains the noninvariant contribution with an ambiguous sign, which contradicts the prediction that $J_s^{\text{chi}}=0$ in the presence of the $\mathbb{Z}_2$ symmetry. Therefore, this discontinuity corresponds to a quantum anomaly of Cooper pairs somewhat similar to the mirror anomaly proposed by Burkov \cite{Burkov18PRL} but associated with the $\mathbb{Z}_2$ symmetry of Eq.~\eqref{eq:symmetry}. Note that the quantum anomaly is restricted to a specific parameter regime (i.e., for $\phi=\pi$, $T=0$ and $L\gg1/\Delta$).

\textit{Anomalous Fraunhofer pattern.\textendash }In the Weyl Josephson junction, the magnetic field $B_{x}$ applied to the N region also modulates the phase difference spatially \cite{Tinkham96BookSC}. The total supercurrent through the junction is then given by
\begin{equation}
I_{s}=\dfrac{W\Phi_{0}}{2\pi\Phi}\int_{-\pi\Phi/\Phi_{0}+\gamma_{0}}^{\pi\Phi/\Phi_{0}+\gamma_{0}}d\phi J_{\text{\ensuremath{s}}}^{\text{tot}}(\phi).\label{eq:Supercurrent}
\end{equation}
Here, $J_{\text{\ensuremath{s}}}^{\text{tot}}(\phi)$ is taking into account the modification by the Zeeman field;  $\gamma_{0}$ is the phase difference at $y=0$; $\Phi=B_{x}WL$ the magnetic
flux threading the N region;  $W$ the junction width; $\Phi_0$ the flux quantum. Plotting the maximum supercurrent $I_{c}=\text{max}[I_{s}(\gamma_{0})]$ as a function of \textbf{$B_{x}$ }or $\Phi$ yields the Fraunhofer pattern \cite{Tinkham96BookSC}.

\begin{figure}[h]
\centering

\includegraphics[width=8cm]{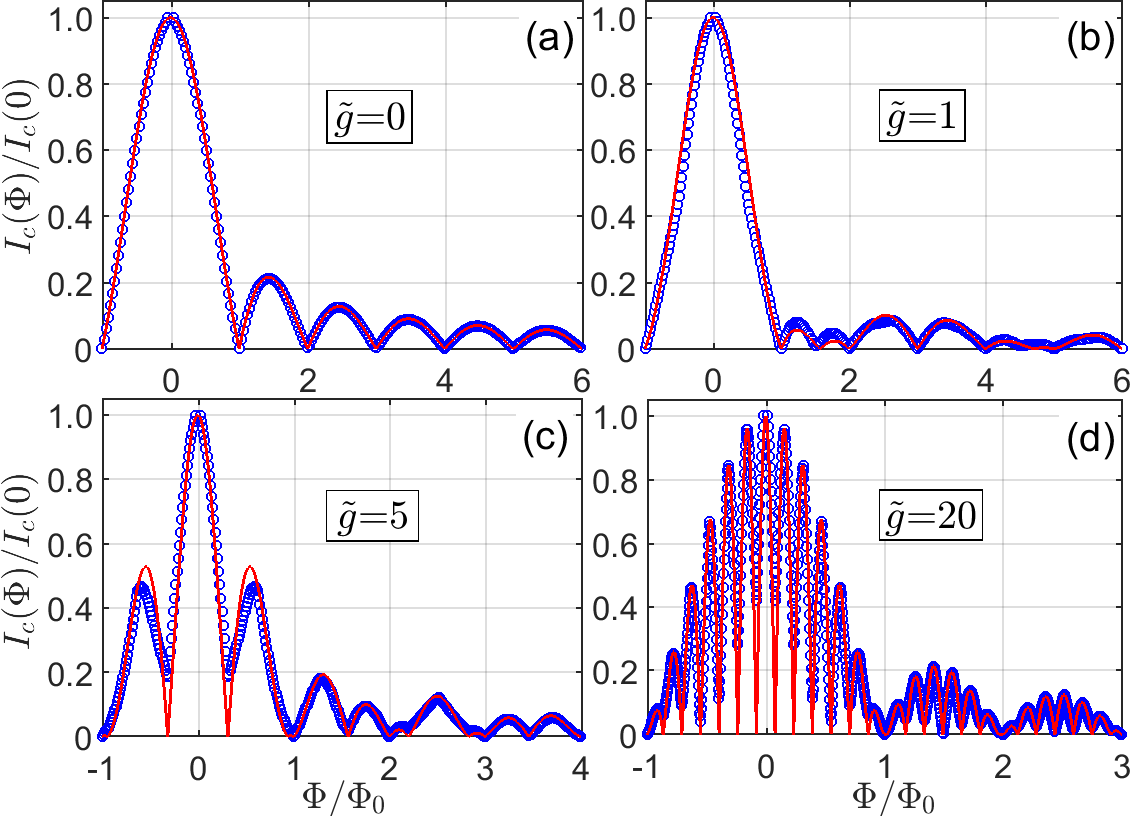}

\caption{Fraunhofer patterns for $\tilde{g}=0,1,5$, and $20$, respectively. The blue circled curves are calculated from Eq.\ (\ref{eq:Supercurrent}), while the red curves are plots of Eq.\ (S5.1) in Ref.~\cite{Supplementary}. $L=1/(100\Delta)$, $\mu_{S}=\mu_{N}=100\Delta$ and $k_{B}T=0.01\Delta$ in all plots. }

\label{Fig:Fraunhofer-pattern}
\end{figure}

With the previously obtained CPRs plugged into Eq.\ (\ref{eq:Supercurrent}), the Fraunhofer patterns are readily calculated and displayed in Fig.\ \ref{Fig:Fraunhofer-pattern}. The Fraunhofer pattern strongly depends on the quantity $\tilde{g}\equiv g\mu_{B}\Phi_{0}/W$. When $\tilde{g}$ is negligible, $\tilde{g}\ll1$, such as in a wide junction with a small $g$-factor, we have a conventional Fraunhofer
pattern shape, i.e., $I_{c}(\Phi)/I_{c}(0)=(\Phi_{0}/\Phi)|\sin(\pi\Phi/\Phi_{0})|$ {[}Fig.\ \ref{Fig:Fraunhofer-pattern}(a){]}. For $\tilde{g}\apprge1$, more (local) maxima appear and their values do not decay monotonically with increasing $\Phi$ or $B_{x}$ anymore {[}Fig.\ \ref{Fig:Fraunhofer-pattern}(b, c, d){]}. For a large $\tilde{g}\apprge10$ such as in a narrow junction with a large $g$-factor, the Fraunhofer pattern exhibits a beating behavior with two frequencies {[}Fig.\ \ref{Fig:Fraunhofer-pattern}(d){]}. We note that these anomalous Fraunhofer patterns, originating from the interference of the supercurrents of different chirality, are a manifestation of the chirality-dependent phase shifts in the CPRs due to the Zeeman field.

\textit{Discussion and summary.\textendash }We briefly discuss the applicability of our results to experiments. In the candidate material of compressively strained HgTe, $g\simeq22.5$ \cite{Madelung-book}. Consider a typical Fermi velocity of $v_{F}\simeq10^{5}\text{ m/s}$. Then, at a magnetic field $B_{x}\text{=}10$ mT, the position shift of Weyl nodes is $\delta k_{z}\simeq10^{-4}\text{nm}^{-1}$ which is one or two orders of magnitude smaller than the separation of Weyl nodes \cite{Ruan16Ncomms}. For junctions with lengths larger than $500$ nm, we could hence obtain an observable phase shift $\phi_{0}\apprge0.1$. In order to have $\tilde{g}\apprge1$, a junction width smaller than $40$ nm is needed in this material.

In summary, we find that in Josephson junctions based inversion-asymmetric WSMs, a perpendicular Zeeman field can realize chirality-dependent $\phi_{0}$-junctions and gives rise to a finite CJC which occurs even without a phase difference. This can be understood by the distinct spin textures near Weyl nodes of opposite chirality. This effect  shows a novel quantum anomaly of Goldstone bosons in a specific parameter regime and manifests itself as an anomalous Fraunhofer pattern.

\textit{Acknowledgments.\textendash }We would like to thank J. Bardarson, D. Breunig,  P. Burset, A. Cappelli, F. Dominguez, C. Fleckenstein, H. Hansson, M. Stehno, G. Tang, N. Traverso Ziani, S. Upadhyay, and X. Wu for useful discussions. This work was supported by the DFG (SPP1666 and SFB1170 \textquotedbl{}ToCoTronics\textquotedbl{}) and the ENB Graduate School on \textquotedbl{}Topological Insulators\textquotedbl{}.


%

\onecolumngrid
\appendix
\clearpage
\newpage

\renewcommand{\thesection}{S\arabic{section}}   
\renewcommand{\thetable}{S\arabic{table}}   
\renewcommand{\thefigure}{S\arabic{figure}}
\numberwithin{equation}{section}
\renewcommand{\thepage}{S\arabic{page}}

\begin{center}
\bf{\large Supplemental Material}
\end{center}

\section{Effective Zeeman term\label{sec:Derivation-of-the}}

Performing the unitary transformation $c_{\uparrow(\downarrow)}^{(\sigma)}=(c_{\sigma,\uparrow}\pm c_{\sigma,\downarrow})/\sqrt{2}$,
$\sigma\in\{A,B\}$, the Zeeman interaction transforms into
\begin{align}
\tilde{H}_{\text{Z}} & =\begin{pmatrix}h_{x} & h_{z}+ih_{y} & 0 & 0\\
h_{z}-ih_{y} & -h_{x} & 0 & 0\\
0 & 0 & h_{x} & h_{z}+ih_{y}\\
0 & 0 & h_{z}-ih_{y} & -h_{x}
\end{pmatrix}\label{eq:Zeman}
\end{align}
in the basis $(c_{\uparrow}^{(A)},c_{\downarrow}^{(A)},c_{\uparrow}^{(B)},c_{\downarrow}^{(B)})$.
The Zeeman field couples spins within the same orbital. The $x$-component
$h_{x}$ is diagonal, whereas the $y$- and $z$-components $h_{y}$,
$h_{z}$ are off-diagonal in this basis. One can observe that $h_{y}$
and $h_{z}$ couple electrons from different Weyl nodes, i.e., Weyl
nodes 1 and 3 to Weyl nodes 2 and 4, whereas $h_{x}$ couples electrons
not only from different Weyl nodes, i.e., Weyl node 1 to Weyl node
3 and Weyl node 2 to Weyl node 4, but also acts within each Weyl node.
At low energies, the basis wave functions $\Psi_{\gamma}({\bf r})=(\psi_{\gamma,\uparrow}({\bf r}),\psi_{\gamma,\downarrow}({\bf r}))$,
$\gamma\in\{1,2,3,4\}$, for the Weyl nodes can be found as
\begin{align}
\Psi_{\gamma}({\bf r}) & =e^{i({\bf k}+{\bf Q}_{\gamma})\cdot{\bf r}}(c_{\uparrow}^{(B)},c_{\downarrow}^{(A)}),\ \ \ \gamma=1,3,\nonumber \\
\Psi_{\gamma}({\bf r}) & =e^{i({\bf k}+{\bf Q}_{\gamma})\cdot{\bf r}}(c_{\uparrow}^{(A)},c_{\downarrow}^{(B)}),\ \ \ \gamma=2,4,\label{eq:Weylbasis2}
\end{align}
where ${\bf Q}_{\gamma}$ is the position of the Weyl node labeled
by $\gamma$ in momentum space. The full spinor basis containing 8
components can be written as
\begin{equation}
\Psi({\bf r})=(\Psi_{1}({\bf r}),\Psi_{2}({\bf r}),\Psi_{3}({\bf r}),\Psi_{4}({\bf r})).\label{eq:NabmuSpinor}
\end{equation}
The projection of the Zeeman interaction (\ref{eq:Zeman}) onto the
spinor (\ref{eq:NabmuSpinor}) results in
\begin{equation}
H_{ij}^{\text{Z}}=\dfrac{1}{\Omega}\int d{\bf r}\psi_{i}^{*}({\bf r})\tilde{H}_{\text{Z}}\psi_{j}({\bf r}),
\end{equation}
where $\psi_{i}$ is the $i$-th component of the spinor (\ref{eq:NabmuSpinor}),
$\Omega=L_{x}L_{y}L_{z}$ with $L_{x,y,z}$ the lengths of the system
along three principal directions, respectively. For example, $H_{1,1}^{\text{Z}}$,
$H_{1,4}^{\text{Z}}$, and $H_{1,5}^{\text{Z}}$ are explicitly given by
\begin{align}
H_{1,1}^{\text{Z}} & =\dfrac{1}{\Omega}\int d^{3}{\bf r}e^{-i{\bf Q}_{1}\cdot{\bf r}}\left(0,0,1,0\right)\tilde{H}_{\text{Z}}e^{i{\bf Q}_{1}\cdot{\bf r}}\left(0,0,1,0\right)^{T}=h_{x},\nonumber \\
H_{1,4}^{\text{Z}} & =\dfrac{1}{\Omega}\int d^{3}{\bf r}e^{-i{\bf Q}_{1}\cdot{\bf r}}\left(0,0,1,0\right)\tilde{H}_{\text{Z}}e^{i{\bf Q}_{2}\cdot{\bf r}}\left(0,0,0,1\right)^{T}=\left(h_{z}+ih_{y}\right)\dfrac{\sin(\beta L_{x})}{\beta L_{x}}\dfrac{\sin(k_{0}L_{z})}{k_{0}L_{z}},\nonumber \\
H_{1,5}^{\text{Z}} & =\dfrac{1}{\Omega}\int d^{3}{\bf r}e^{-i{\bf Q}_{1}\cdot{\bf r}}\left(0,0,1,0\right)\tilde{H}_{\text{Z}}e^{i{\bf Q}_{3}\cdot{\bf r}}\left(0,0,1,0\right)^{T}=h_{x}\dfrac{\sin(k_{0}L_{z})}{k_{0}L_{z}}.
\end{align}
Suppose the lengths $L_{x,z}$ or separations of Weyl nodes $k_{0}\ \text{and}\ \beta$
are large enough such that $\beta L_{x}\gg1$ and $k_{0}L_{z}\gg1.$
Then, inter-Weyl node couplings, such as $H_{1,4}^{\text{Z}}$ and
$H_{1,5}^{\text{Z}}$, vanish. Therefore, the projection of the Zeeman
interaction becomes diagonal in the Weyl-node space and gives rise to Eq.\ (3) in the Letter for each Weyl node.

\section{Josephson currents\label{sec:Scattering-and-interface}}

In this section, we calculate the scattering matrix in the N region, the reflection matrix at the interfaces, and finally the Josephson
currents.

\subsection{Scattering and interface reflection matrices}

We take the positive chirality sector for illustration and consider
the Hamiltonian given in the main text and below explicitly
\begin{align}
h_{\text{BdG}}^{+} & =\begin{pmatrix}h_{\text{BdG}} & \cancel{0}\\
\cancel{0} & h_{\text{BdG}}
\end{pmatrix},\label{eq:BigBdG}\\
h_{\text{BdG}} & =\begin{pmatrix}-i\partial_{{\bf r}}\cdot\bm{s}+h({\bf r})s_{z}-\mu({\bf r})s_{0} & \Delta_{s}({\bf r})e^{i\text{sgn}(z)\phi/2}s_{0}\\
\Delta_{s}({\bf r})e^{-i\text{sgn}(z)\phi/2}s_{0} & i\partial_{{\bf r}}\cdot\bm{s}+h({\bf r})s_{z}+\mu({\bf r})s_{0}
\end{pmatrix},\label{eq:BdG-Hamiltonian}
\end{align}
where $\cancel{0}$ is the $4\times4$ null matrix, $\mu({\bf r})$,
$h({\bf r})$, and $\Delta_{s}({\bf r})$ are spatially dependent
quantities, as described in the main text. The Hamiltonian decouples
into two identical blocks given by Eq.\ (\ref{eq:BdG-Hamiltonian}).
In this section, we deal with Eq.\ (\ref{eq:BdG-Hamiltonian}), which
is enough for the transport problem. In the N region, the basis functions
at a given energy $\varepsilon$ can be written as (the factor $e^{ik_{x}x+ik_{y}y}$ is omitted for simplicity)
\begin{align}
\varphi_{e+}(z) & =(\cos\alpha_{e},e^{i\theta_{{\bf k}}}\sin\alpha_{e},0,0)^{T}e^{ik_{e+}z},\ \ \ \ \ \varphi_{e-}(z)=(e^{-i\theta_{{\bf k}}}\sin\alpha_{e},\cos\alpha_{e},0,0)^{T}e^{ik_{e-}z},\nonumber \\
\varphi_{h+}(z) & =(0,0,\cos\alpha_{h},-e^{i\theta_{{\bf k}}}\sin\alpha_{h})^{T}e^{ik_{h-}z},\ \ \varphi_{h-}(z)=(0,0,-e^{-i\theta_{{\bf k}}}\sin\alpha_{h},\cos\alpha_{h})^{T}e^{ik_{h+}z},
\end{align}
where $\theta_{{\bf k}}=\arctan(k_{y}/k_{x})$, $\alpha_{e(h)}=\arctan(k_{\parallel}/k_{e(h)})/2$,
$k_{\parallel}=(k_{x}^{2}+k_{y}^{2})^{1/2}$, $k_{e\pm}=-h_{x}\pm k_{e}$,
$k_{h\pm}=h_{x}\pm k_{h}$ and $k_{e(h)}=[(\varepsilon\pm\mu_{N})^{2}-k_{\parallel}^{2}]^{1/2}$.
$T$ means the transpose of a matrix. In the two S regions, the basis
functions are
\begin{align}
\varphi_{qe+}(z)= & (e^{i\beta}\cos\tilde{\alpha}_{e},e^{i\beta}e^{i\theta_{{\bf k}}}\sin\tilde{\alpha}_{e},e^{-i\phi_{t}}\cos\tilde{\alpha}_{e},e^{-i\phi_{t}}e^{i\theta_{{\bf k}}}\sin\tilde{\alpha}_{e})^{T}e^{i\tilde{k}_{e}z},\nonumber \\
\varphi_{qe-}(z)= & (e^{i\beta}e^{-i\theta_{{\bf k}}}\sin\tilde{\alpha}_{e},e^{i\beta}\cos\tilde{\alpha}_{e},e^{-i\phi_{t}}e^{-i\theta_{{\bf k}}}\sin\tilde{\alpha}_{e},e^{-i\phi_{t}}\cos\tilde{\alpha}_{e})^{T}e^{-i\tilde{k}_{e}z},\nonumber \\
\varphi_{qh+}(z)= & (e^{i\phi_{t}}\sin\tilde{\alpha}_{h},e^{i\phi_{t}}e^{i\theta_{{\bf k}}}\cos\tilde{\alpha}_{h},e^{i\beta}\sin\tilde{\alpha}_{h},e^{i\beta}e^{i\theta_{{\bf k}}}\cos\tilde{\alpha}_{h})^{T}e^{-i\tilde{k}_{h}z},\nonumber \\
\varphi_{qh-}(z)= & (e^{i\phi_{t}}e^{-i\theta_{{\bf k}}}\cos\tilde{\alpha}_{h},e^{i\phi_{t}}\sin\tilde{\alpha}_{h},e^{i\beta}e^{-i\theta_{{\bf k}}}\cos\tilde{\alpha}_{h},e^{i\beta}\sin\tilde{\alpha}_{h})^{T}e^{i\tilde{k}_{h}z},
\end{align}
where $t\in\{r,l\}$ labels the pairing phases $\phi_{l}=-\phi/2$
and $\phi_{r}=\phi/2$ on the left and right hand sides, respectively;
$\tilde{\alpha}_{e(h)}=\arctan(k_{\parallel}/\tilde{k}_{e(h)})/2$
and $\tilde{k}_{e(h)}=[\left(\mu_{S}\pm\Omega\right)^{2}-k_{\parallel}^{2}]^{1/2}$.
For subgap energies $\varepsilon\leqslant\Delta$, $\beta=\arccos(\varepsilon/\Delta)$,
and $\Omega=i(\Delta^{2}-\varepsilon^{2})^{1/2}$, while for supragap
energies $\varepsilon>\Delta$, $\beta=-i\text{arccosh}(\varepsilon/\Delta)$,
and $\Omega=\text{sgn}(\varepsilon)(\varepsilon^{2}-\Delta^{2})^{1/2}$.
Therefore, the wave function can be expanded in terms of these basis
functions as
\begin{equation}
\Psi(z)=\begin{cases}
N_{e+}\varphi_{e+}(z)+N_{e-}\varphi_{e-}(z)+N_{h+}\varphi_{h+}(z)+N_{h-}\varphi_{h-}(z), & |z|<L/2,\\
S_{e+}\varphi_{qe+}(z)+S_{e-}\varphi_{qe-}(z)+S_{h+}\varphi_{qh+}(z)+S_{h-}\varphi_{qh-}(z), & |z|>L/2,
\end{cases}\label{eq:Wave-function}
\end{equation}
where superposition coefficients $N_{e(h)\pm}$ and $S_{e(h)\pm}$
are determined by the boundary conditions.

The reflection matrix is then determined by matching the wave function in Eq.~(\ref{eq:Wave-function}) at the interfaces at $z=\pm L/2$ and found
explicitly as
\begin{equation}
\mathcal{R}_{A}^{+}=\begin{pmatrix}R_{ee}e^{-i\theta_{{\bf k}}} & R_{eh}e^{-i\phi/2} & 0 & 0\\
R_{he}e^{i\phi/2} & R_{hh}e^{i\theta_{{\bf k}}} & 0 & 0\\
0 & 0 & R_{ee}e^{i\theta_{{\bf k}}} & R_{eh}e^{i\phi/2}\\
0 & 0 & R_{he}e^{-i\phi/2} & R_{hh}e^{-i\theta_{{\bf k}}}
\end{pmatrix},\label{eq:ReflectionMatrix}
\end{equation}
where
\begin{align}
R_{ee}= & -[e^{i\beta}\sin(\alpha_{e}-\tilde{\alpha}_{e})\sin(\alpha_{h}+\tilde{\alpha}_{h})-e^{-i\beta}\sin(\alpha_{h}+\tilde{\alpha}_{e})\sin(\alpha_{e}-\tilde{\alpha}_{h})]/\mathcal{Z},\nonumber \\
R_{eh}= & -\cos(2\alpha_{h})\sin(\tilde{\alpha}_{e}-\tilde{\alpha}_{h})/\mathcal{Z},\ \ R_{he}=-\cos(2\alpha_{e})\sin(\tilde{\alpha}_{e}-\tilde{\alpha}_{h})/\mathcal{Z},\nonumber \\
R_{hh}= & [e^{i\beta}\cos(\alpha_{e}+\tilde{\alpha}_{e})\cos(\alpha_{h}-\tilde{\alpha}_{h})-e^{-i\beta}\cos(\alpha_{e}+\tilde{\alpha}_{h})\cos(\alpha_{h}-\tilde{\alpha}_{e})]/\mathcal{Z},
\end{align}
with $\mathcal{Z}=e^{i\beta}\cos(\alpha_{e}+\tilde{\alpha}_{e})\sin(\alpha_{h}+\tilde{\alpha}_{h})-e^{-i\beta}\cos(\alpha_{e}+\tilde{\alpha}_{h})\sin(\alpha_{h}+\tilde{\alpha}_{e})$
\cite{Songbo18PRB1}. The reflection matrix is block-diagonal with
two blocks describing the reflection property of the two interfaces,
respectively. Assume that the N region is clean enough such that no
scattering occurs between different modes. The scattering matrix is
then given by the dynamic phases accumulated as the particles and
holes propagate along the N region:
\begin{equation}
\mathcal{S}_{N}^{+}=\begin{pmatrix}0 & 0 & e^{-ik_{e-}L} & 0\\
0 & 0 & 0 & e^{-ik_{h-}L}\\
e^{ik_{e+}L} & 0 & 0 & 0\\
0 & e^{ik_{h+}L} & 0 & 0
\end{pmatrix}.\label{eq:ScatteringMatrix}
\end{equation}
The Zeeman field $h_{x}$ does not alter the basis states but only
enters the wave vectors $k_{e\pm}$ and $k_{h\pm}$ which are canceled
out on matching the wave function. Therefore, $h_{x}$ does not affect
the reflection $\mathcal{R}_{A}^{+}$ but scattering matrix $\mathcal{S}_{N}^{+}$,

\subsection{Josephson current}

We apply the method described in Refs.\ \cite{Brouwer97SSC1,Dolcini07PRB1,Beenakker13PRL1}
to calculate the Josephson current densities
\begin{equation}
J_{s}^{\pm}(\phi)=-\dfrac{16ek_{B}T}{\hbar}\sum_{\bf{k}_{\parallel}}\sum_{n=0}^{\infty}\dfrac{d}{d\phi}\ln D_{\pm}({\bf{k}_{\parallel}},\phi;i\omega_{n}),\label{eq:Josephson_current}
\end{equation}
where the sums run over all transverse momenta $\bf{k}_{\parallel}$ and fermionic Matsubara frequencies $\omega_{n}=(2n+1)\pi k_{B}T$, $-e$ is the electron charge, $T$ the temperature, $k_{B}$ the Boltzmann constant and $D_{\pm}=\det(1-\mathcal{R}_{A}^{\pm}\mathcal{S}_{N}^{\pm}).$
The index $+$($-$) denotes positive(negative) chirality. This method takes into account both contributions from discrete Andreev bound states and the continuum spectrum. The condition $D_{\pm}(\bf{k}_{\parallel},\phi;\varepsilon)=0$ determines the energy spectrum of Andreev bound states with energy within the superconducting gap.

With Eqs.\ (\ref{eq:ReflectionMatrix}) and (\ref{eq:ScatteringMatrix}),
$D_{+}$ is obtained as
\begin{align}
D_{+}= & 1-R_{ee}^{2}e^{2ik_{e}L}-R_{hh}^{2}e^{2ik_{h}L}+(R_{ee}R_{hh}-R_{eh}R_{he})^{2}e^{2i(k_{e}+k_{h})L}-2R_{eh}R_{he}\cos\phi'e^{i(k_{e}+k_{h})L}.
\end{align}
The phase dependence in $D_{+}$ comes from the finite Andreev reflection
at the interfaces. The Zeeman field $h_{x}$ leads directly to a shift
in the phase difference $\phi'=\phi+\phi_{0}$ with $\phi_{0}=2h_{x}L$.
For $\mu_{N}=\mu_{S}\gg\Delta$, $R_{ee}\approx R_{hh}\approx0$ and
$R_{eh}\approx R_{he}\approx e^{-i\beta}$.

\subsection{Josephson current in the long junction and zero temperature limit}

In the long junction and zero temperature limit, only low-energy Andreev bound states with energy $|\epsilon|\ll\Delta$ are relevant for the Josephson current \cite{Bardeen72PRB1}. For simplicity, we put $\Delta\rightarrow\infty$ so that the low-energy states are well localized in the N region. Andreev bound states are determined by the condition $D_{\pm}=0$. But here, we can make use of the long junction assumption and find the Andreev
bound states by using the method of Ref.\ \cite{Bardeen72PRB1}. For a given transverse momentum ${\bf k}_{\parallel}$, there are two Fermi points $\pm k_{F}$ satisfying $(k_{F}^{2}+k_{\parallel}^{2})^{1/2}=\mu_{N}$ (we assume $\mu_{N}>0$ without loss of generality). If we take $k_{z}=\pm k_{F}+q_{z}$ with $q_{z}$ small, then the low-energy excitations measured from $\mu_{N}$ are described by a helical model
\begin{equation}
H_{1}= v_{F}q_{z}\sigma_z',
\end{equation}
where $v_{F}=(\mu_{N}^{2}-k_{\parallel}^{2})^{1/2}/\mu_{N}$ is the effective Fermi velocity and the Pauli matrix $\sigma_z'$ acts on the two chiral bands. For the positive and negative chirality sectors, the allowed values of $q_{z}$ that satisfy the boundary conditions \cite{Crepin14PRL}
at $L=\pm L/2$ are found as
\begin{equation}
q_{n}^{\pm}=\dfrac{\pi}{L}\left(n+\dfrac{1}{2}+\dfrac{\phi\pm\phi_{0}}{2\pi}\right),\ \ \ n\in\mathbb{Z},
\end{equation}
respectively. Both the phase difference and Zeeman field have been taken into account. Therefore, the energies of Andreev bound
states are
\begin{equation}
\epsilon_{n,\eta}^{\pm}=\eta\dfrac{\pi v_{F}}{L}\left(n+\dfrac{1}{2}+\dfrac{\phi\pm\phi_{0}}{2\pi}\right),\ \ \ \eta\in\{-1,1\}.\label{eq:LLL}
\end{equation}
Note that for $k_{\parallel}=0$, $\epsilon_{n}$ can
also be found easily by solving $D_{\pm}=0$ with $\epsilon/\Delta\rightarrow0$. The result is in agreement with Eq.\ (\ref{eq:LLL}). These energy levels are linear in $\phi\pm\phi_{0}$ and cross at $\phi\pm\phi_{0}=\pi$ (and other quantized values $\phi\pm\phi_{0}=(2n+1)\pi$ with $n\in\mathbb{Z}$). The level crossings at $\phi\pm\phi_{0}=(2n+1)\pi$ are protected by a pseudo time-reversal symmetry (with the operator $i\mathcal{K}\sigma_y'$, $\mathcal{K}$ is the complex conjugation).

\begin{figure}[H]
\centering

\includegraphics[height=5cm]{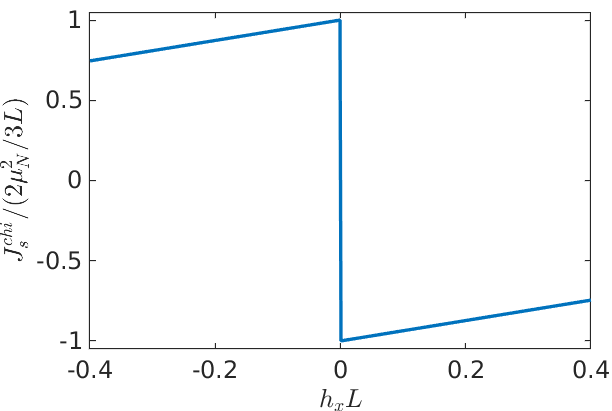}

\caption{Chirality Josephson current $J_{s}^{\text{chi}}$ at $\phi=\pi$ and
$T=0$ as a function of $h_{x}L$ in a long Josephson junction $L\gg1/\Delta$. }

\label{Fig:DiscontinuousJump}
\end{figure}

The Josephson currents of each chirality can be related to the eigenenergies
by \cite{Beenakker91PRL1,Bagwell92PRB1}
\begin{align}
J_{s}^{\pm} & =\sum_{{\bf k}_{\parallel}}j_{s}^{\pm},\ \ j_{s}^{\pm}=\dfrac{e}{\hbar}\dfrac{\partial F_{\pm}(\phi)}{\partial\phi},\ \ F_{\pm}(\phi)=2\sum_{n}\epsilon_{n,\eta}^{\pm}f_{F}(\epsilon_{n,\eta}^{\pm}),\label{eq:JosephsonCurrent-formula-1}
\end{align}
where $F_{\pm}$ is the free energy and $f_{F}(E)$ is the Fermi distribution function. At zero temperature, $f_{F}(E)=\Theta(-E)$. Due to
the linearized spectrum, there are infinitely many negative energy
states forming the Dirac sea. Nevertheless, $J_{s}^{\pm}(\phi)$
can be calculated by the method of Bosonization \cite{Crepin16PE1}
or ultraviolet regularization \cite{Sticlet13PRB1,Manton85AP1} since
the energy states deep in the Dirac sea have exponentially small contribution
to the physical properties of the system. For $-\pi<\phi\pm\phi_{0}<3\pi$,
$j_{s}^{\pm}$ is found as
\begin{align}
j_{s}^{\pm} & =\dfrac{2\pi v_{F}}{L}\Big[\dfrac{\phi\pm\phi_{0}}{\pi}-\text{sgn}(\phi\pm\phi_{0}-\pi)-1\Big].\label{eq:JC_anomaly}
\end{align}
For each ${\bf k}_{\parallel}$, the supercurrent has a piecewise
phase-dependence with a discontinuous jump at $\phi\pm\phi_{0}=\pi$.
Summing over ${\bf k}_{\parallel}$ and taking the difference of $J_{s}^{\pm}$,
we obtain the chirality Josephson current. At $\phi=\pi$, the chirality
Josephson current $J_{s}^{\text{chi}}$ for small $h_{x}L$ is given by
\begin{align}
J_{s}^{\text{chi}} & =\dfrac{2\mu_{N}^{2}}{3L}\Big[\dfrac{2h_{x}L}{\pi}-\text{sgn}(h_{x})\Big],\ \ \ \ |h_{x}L|<\pi.\label{eq:JC_anomaly-1-1}
\end{align}
It shows a discontinuous jump at $h_{x}=0$ (see Fig.\ \ref{Fig:DiscontinuousJump}).

The discontinuity in Eq.\ \eqref{eq:JC_anomaly-1-1} is robust against a non-magnetic and perturbative impurity barrier modeled by a scalar potential $U({\bf{r}})=\delta U [\Theta(z-z_1)-\Theta(z-z_2)]$ with $-L/2\leqslant z_1<z_2\leqslant L/2$. For simplicity, we assume a uniform barrier potential $U({\bf{r}})$ in $x$ and $y$ directions. Then, the transverse momentum $\bf{k}_{\parallel}$ is preserved. This potential may account for an interface barrier or smooth (particularly in $x$ and $y$ directions) disorder in the N region. If  $\delta U$  is small enough such that the linearized model remains valid, then, the scalar potential only generates phase shifts for the two chiral particles without coupling them.  By imposing the boundary conditions, the energies of Andreev bound states are found as
\begin{equation}
\epsilon_{n,\eta}^{\pm}=\eta\dfrac{\pi v_{F}}{L}\left(n+\dfrac{1}{2}+\dfrac{\phi\pm\phi_{0}}{2\pi}\right)+U_0,\ \ \ \eta\in\{-1,1\},\label{eq:LLL1}
\end{equation}
where $U_0=(z_2-z_1)\delta U/L$. The only effect of the impurity barrier is to shift the spectrum by a global constant $U_0$ which does not depend on the phase difference. The energy crossings at $\phi\pm\phi_{0}=\pi$ of the clean spectrum in Eq.\ \eqref{eq:LLL}  are thus preserved since the scalar potential does not break the underlying pseudo time-reversal symmetry. This behavior is a manifestation of the absence of backscattering or Klein tunneling \cite{Katsnelson06nphys1}  in a helical liquid. Plugging Eq.~\eqref{eq:LLL1} into Eq.~\eqref{eq:JosephsonCurrent-formula-1}, we arrive at exactly the same formulas as Eqs.~\eqref{eq:JC_anomaly} and \eqref{eq:JC_anomaly-1-1}. Note that, for magnetic or strong impurities \cite{Bagwell92PRB1}, backscattering is no longer prohibited. It opens gaps between the energy levels of the two chiral particles. Then, the discontinuity in Eq.~\eqref{eq:JC_anomaly-1-1} will be smoothed out.

\section{Normal-state resistance of the junction\label{sec:Normal-state-resistance-of}}

In this section, we apply the scattering matrix formalism \cite{Blonder82PRB1}
to calculate the normal-state resistance of the junction. The wave
function of a scattering state in which an electron is incident on
the left-hand side and moves to the right is given by
\begin{equation}
\Psi(z)=\begin{cases}
\bar{\varphi}_{\overrightarrow{e}}(z)+r_{0}\bar{\varphi}_{\overleftarrow{e}}(z), & z<-L/2,\\
\alpha_{0}\varphi_{\overrightarrow{e}}(z)+\beta_{0}\varphi_{\overleftarrow{e}}(z), & |z|<L/2,\\
t_{0}\bar{\varphi}_{\overrightarrow{e}}(z), & z>L/2.
\end{cases}
\end{equation}
Consider the scattering state at the Fermi level. Then, the basis
functions are given by
\begin{align}
\varphi_{\overrightarrow{e}}(z)= & (\cos(\alpha/2),\sin(\alpha/2))^{T}e^{ik_{e}z},\ \ \varphi_{\overleftarrow{e}}(z)=(\sin(\alpha/2),\cos(\alpha/2))^{T}e^{-ik_{e}z},\nonumber \\
\bar{\varphi}_{\overrightarrow{e}}(z)= & (\cos(\bar{\alpha}/2),\sin(\bar{\alpha}/2))^{T}e^{i\bar{k}_{e}z},\ \ \bar{\varphi}_{\overleftarrow{e}}(z)=(\sin(\bar{\alpha}/2),\cos(\bar{\alpha}/2))^{T}e^{-i\bar{k}_{e}z},
\end{align}
where the angles $\alpha$ and $\bar{\alpha}$ are defined by $\sin\alpha=k_{\parallel}/\mu_{N}$,
$\cos\alpha=k_{e}/\mu_{N}$, $\sin\bar{\alpha}=k_{\parallel}/\mu_{S}$
and $\cos\bar{\alpha}=\bar{k}_{e}/\mu_{S}$, respectively. Note that
the scattering state exists only when $\bar{k}_{e}=(\mu_{S}^{2}-k_{\parallel}^{2})^{1/2}$
is real. The coefficients $\alpha_{0},$ $\beta_{0},$ $t_{0}$ and
$r_{0}$ are determined by requiring the continuity of $\Psi(z)$
at the interfaces at $z=\pm L/2$. The transmission probability $T_{0}\equiv|t_{0}|^{2}$
of current density across the junction can be found as
\begin{align}
T_{\text{0}} & =\left|\dfrac{\cos\alpha\cos\bar{\alpha}}{\cos\alpha\cos\bar{\alpha}\cos(k_{e}L)+i\left(\sin\alpha\sin\bar{\alpha}-1\right)\sin(k_{e}L)}\right|^{2}.
\end{align}
This implies that $T_{0}$ is an oscillatory function of the junction
length $L$ for $\mu_{N}>\mu_{S}$. For the short junction limit with
$L=0$, a uniform chemical potential $\mu_{N}=\mu_{S},$ or normal
incidence $k_{\parallel}=0$, we obtain the perfect transmission $T_{\text{0}}=1$.

The resistance of the junction at zero temperature can then be calculated
by the Landauer formula
\begin{equation}
R_{n}^{-1}=\dfrac{4e^{2}}{h}\sideset{}{_{{\bf {\bf k}_{\parallel}}}}\sum T_{0}({\bf k}_{\parallel}),
\end{equation}
where the factor $4$ is due to four Weyl cones in the system. For
$\mu_{N}=\mu_{S}$ or $L=0$, $R_{n}=(h/e^{2})(\pi/\mu_{S}^{2})$.

\section{Symmetry analysis\label{sec:Symmetry-analysis}}

At low energies, the Hamiltonian for the two Weyl nodes of positive
chirality can be written as
\begin{align*}
H_{+}({\bf r}) & =-i\tau_{0}(\partial_{x}s_{x}+\partial_{y}s_{y}+\partial_{z}s_{z})-\tau_{z}\left(\beta s_{x}+k_{0}s_{z}\right)
\end{align*}
in the basis $\tilde{\Psi}_{+}=(e^{-i{\bf Q}_{1}\cdot{\bf r}}\Psi_{1},e^{-i{\bf Q}_{2}\cdot{\bf r}}\Psi_{2})$.
Here, $s_{i}$ and $\tau_{i}$ ($i=0,x,y,z$) act in spin and Weyl-node
spaces, respectively. For the Weyl nodes of negative chirality, the
Hamiltonian reads
\begin{align*}
H_{-}({\bf r}) & =-i\tau_{0}(\partial_{x}s_{x}+\partial_{y}s_{y}-\partial_{z}s_{z})-\tau_{z}\left(\beta s_{x}+k_{0}s_{z}\right)
\end{align*}
in the basis $\tilde{\Psi}_{-}=(e^{-i{\bf Q}_{3}\cdot{\bf r}}\Psi_{3},e^{-i{\bf Q}_{4}\cdot{\bf r}}\Psi_{4})$.
According to Eq.\ (\ref{eq:Weylbasis2}), the Hamiltonian for opposite
chirality share the same basis $\tilde{\Psi}_{+}=\tilde{\Psi}_{-}=(c_{\uparrow}^{(B)},c_{\downarrow}^{(A)},c_{\uparrow}^{(A)},c_{\downarrow}^{(B)})$.
In this basis, the time-reversal operator reads $\mathcal{T}=-i\tau_{x}s_{y}\mathcal{K},$
where $\mathcal{K}$ denotes the complex conjugation. In addition,
there exists an emergent symmetry operation
\begin{equation}
\mathcal{U}=i\tau_{y}s_{y}\mathcal{R}_{x}=e^{i\pi s_{y}\tau_{y}R_{x}/2},
\end{equation}
where $\mathcal{R}_{x}$ is the reflection operator about the $yz$
plane, (i.e., $x$$\rightarrow$$-x$). In the absence of magnetic
fields, each chirality sector is invariant under time-reversal, i.e.,
$[H_{\pm}({\bf r}),\mathcal{T}]=0$. They are connected to each other
by the emergent symmetry, i.e., $\mathcal{U}H_{+}({\bf r})\mathcal{U}^{-1}=H_{-}({\bf r})$.
In the full basis spinor $(\tilde{\Psi}_{+},\text{\ensuremath{\tilde{\Psi}_{-}}})$,
we can define an operator as
\begin{equation}
\mathcal{U}_{5}=\begin{pmatrix}\cancel{0} & \mathcal{U}\\
\mathcal{U} & \cancel{0}
\end{pmatrix}.
\end{equation}
which commutes with the full Hamiltonian $H=\text{diag}(H_{+}({\bf r}),H_{-}({\bf r}))$,
i.e., $[H,\mathcal{U}_{5}]=0.$ This implies that the system has a
discrete $\mathbb{Z}_{2}$ symmetry which exchanges the two chirality
sectors. We thus call $\mathcal{U}$ the $\mathbb{Z}_{2}$ (exchange)
symmetry.

The BdG Hamiltonians for the two chirality sectors can be written as
\begin{align}
\mathcal{H}_{\text{BdG}}^{\pm}(\phi) & =\begin{pmatrix}H_{\pm}({\bf r})-\mu({\bf r})\tau_{0}s_{0} & i\Delta_{s}({\bf r})e^{i\text{sgn}(z)\phi/2}\tau_{x}s_{y}\\
-i\Delta_{s}({\bf r})e^{-i\text{sgn}(z)\phi/2}\tau_{x}s_{y} & -H_{\pm}^{*}({\bf r})+\mu({\bf r})\tau_{0}s_{0}
\end{pmatrix}\label{eq:Sy-BdG-H}
\end{align}
in the Nambu basis $(c_{\uparrow}^{(B)},c_{\downarrow}^{(A)},c_{\uparrow}^{(A)},c_{\downarrow}^{(B)},c_{\uparrow}^{(B)\dagger},c_{\downarrow}^{(A)\dagger},c_{\uparrow}^{(A)\dagger},$
$c_{\downarrow}^{(B)\dagger})$. Accordingly, the extended operator
of $\mathcal{U}$ in this Nambu basis reads $\mathcal{U}_{\text{BdG}}=\text{diag}\left(\mathcal{U},-\mathcal{U}\right).$
It relates the BdG Hamiltonians as
\begin{equation}
\mathcal{U}_{\text{BdG}}\mathcal{H}_{\text{BdG}}^{+}(\phi)\mathcal{U}_{\text{BdG}}^{-1}=\mathcal{H}_{\text{BdG}}^{-}(\phi).\label{eq:ChiralBdG}
\end{equation}
The BdG equation for positive chirality is described by
\begin{equation}
\mathcal{H}_{\text{BdG}}^{+}(\phi)\psi_{n}^{+}=E_{n}^{+}(\phi)\psi_{n}^{+},\label{eq:BdGequation-2}
\end{equation}
where $\psi_{n}$ and $E_{n}$ are eigenstate and eigenenergy labeled
by an index $n$, respectively. Making use of Eq.\ (\ref{eq:ChiralBdG}),
Eq.\ (\ref{eq:BdGequation-2}) can be transformed to
\begin{equation}
\mathcal{H}_{\text{BdG}}^{-}(\phi)\mathcal{U}_{\text{BdG}}\psi_{n}^{+}=E_{n}^{+}(\phi)\mathcal{U}_{\text{BdG}}\psi_{n}^{+}.\label{eq:BdGequation-1-1}
\end{equation}
This indicates that $\mathcal{H}_{\text{BdG}}^{\pm}(\phi)$ have exactly
the same eigenenergies. Namely,
\begin{equation}
E_{n}^{-}(\phi)=E_{n}^{+}(\phi).\label{eq:relation-1}
\end{equation}
Using the formula for the Josephson currents of each chirality
\begin{align}
J_{s}^{\pm}(\phi) & =\dfrac{e}{\hbar}\dfrac{\partial F_{\pm}(\phi)}{\partial\phi},\ F_{\pm}(\phi)=\sum_{n}E_{n}^{\pm}f_{F}(E_{n}^{\pm}).\label{eq:JosephsonCurrent-formula}
\end{align}
With the help of Eq.\ (\ref{eq:relation-1}), we find that the Josephson
currents for the two chirality sectors have to be equal,
\begin{equation}
J_{s}^{+}(\phi)=J_{s}^{-}(\phi).
\end{equation}

The presence of $h_{x}$ breaks both the symmetries $\mathcal{T}$
and $\mathcal{U}$ simultaneously. However, the system preserves a
combined symmetry defined by the product of $\mathcal{T}$ and $\mathcal{U}$,
\begin{equation}
\mathcal{S}=\mathcal{T}\mathcal{U}=-i\tau_{z}\mathcal{R}_{x}\mathcal{K},
\end{equation}
as represented explicitly by
\begin{align}
\mathcal{S}[H_{+}({\bf r})+h({\bf r})\tau_{0}s_{z}]\mathcal{S}^{-1} & =H_{-}({\bf r})+h({\bf r})\tau_{0}s_{z}.
\end{align}
Thus, $\mathcal{S}$ exchanges the two chirality sectors even in the
presence of $h_{x}$. We call it a magnetic $\mathbb{Z}_{2}$ symmetry.
In the presence of $h_{x},$ the BdG Hamiltonians\ (\ref{eq:Sy-BdG-H})
are modified to
\begin{align}
\mathcal{\tilde{H}}_{\text{BdG}}^{\pm}(\phi) & =\begin{pmatrix}H_{\pm}({\bf r})+h({\bf r})\tau_{0}s_{z}-\mu({\bf r})\tau_{0}s_{0} & i\Delta_{s}({\bf r})e^{i\text{sgn}(z)\phi/2}\tau_{x}s_{y}\\
-i\Delta_{s}({\bf r})e^{-i\text{sgn}(z)\phi/2}\tau_{x}s_{y} & -H_{\pm}^{*}({\bf r})-h({\bf r})\tau_{0}s_{z}-\mu({\bf r})\tau_{0}s_{0}
\end{pmatrix}.
\end{align}
They are related by
\begin{equation}
\mathcal{S}_{\text{BdG}}\mathcal{H}_{\text{BdG}}^{+}(\phi)\mathcal{S}_{\text{BdG}}^{-1}=\mathcal{H}_{\text{BdG}}^{-}(-\phi),\label{eq:MMRS}
\end{equation}
with $\mathcal{S}_{\text{BdG}}=\text{diag (\ensuremath{\mathcal{S}},-\ensuremath{\mathcal{S}})}$
the corresponding extended operator of $\mathcal{S}$ in the Nambu
basis. In this case, we do not have the relation in Eq.\ (\ref{eq:BdGequation-1-1})
but still
\begin{equation}
\mathcal{H}_{\text{BdG}}^{-}(\phi)\mathcal{S}_{\text{BdG}}\psi_{n}^{+}=E_{n}^{+}(-\phi)\mathcal{S}_{\text{BdG}}\psi_{n}^{+}.\label{eq:BdGequation-1}
\end{equation}
Consequently,
\begin{equation}
E_{n}^{-}(\phi)\neq E_{n}^{+}(\phi),\ \ E_{n}^{-}(\phi)=E_{n}^{+}(-\phi).\label{eq:relation}
\end{equation}
Therefore, using Eq.\ (\ref{eq:JosephsonCurrent-formula}), we find
$J_{s}^{+}(\phi)\neq J_{s}^{-}(\phi)$, and
\begin{equation}
J_{s}^{+}(\phi)=-J_{s}^{-}(-\phi).\label{eq:Relation-Jc-chirality}
\end{equation}
The relation in Eq.\ (\ref{eq:Relation-Jc-chirality}) holds even
in the presence of $h_{x}$ which breaks time-reversal symmetry.

\section{Fraunhofer patterns\label{sec:Fraunhofer-patterns}}

\begin{figure}[H]
\centering

\includegraphics[width=16cm]{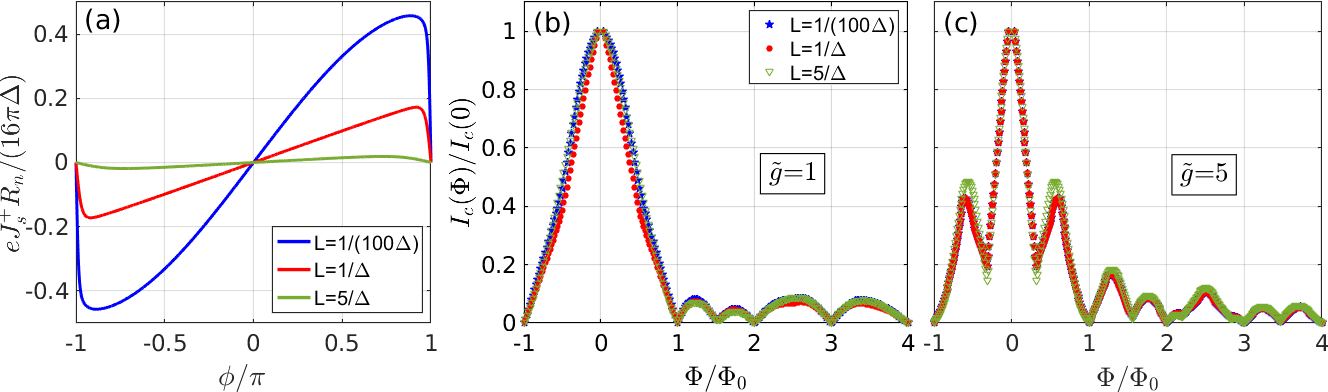}

\caption{Current-phase relations for positive chirality in the absence of magnetic
fields (a) and Fraunhofer patterns for $\tilde{g}=1$ (b) and $5$
(c) in junctions with lengths $L=1/(100\Delta),\ 1/\Delta$, and $5/\Delta$,
respectively. Other parameters are $\mu_{S}=\mu_{N}=100\Delta$ and
$k_{B}T=0.01\Delta$. }

\label{Fig:Fraunhofer-pattern-1}
\end{figure}

\begin{figure}[h]
\centering

\includegraphics[width=16cm]{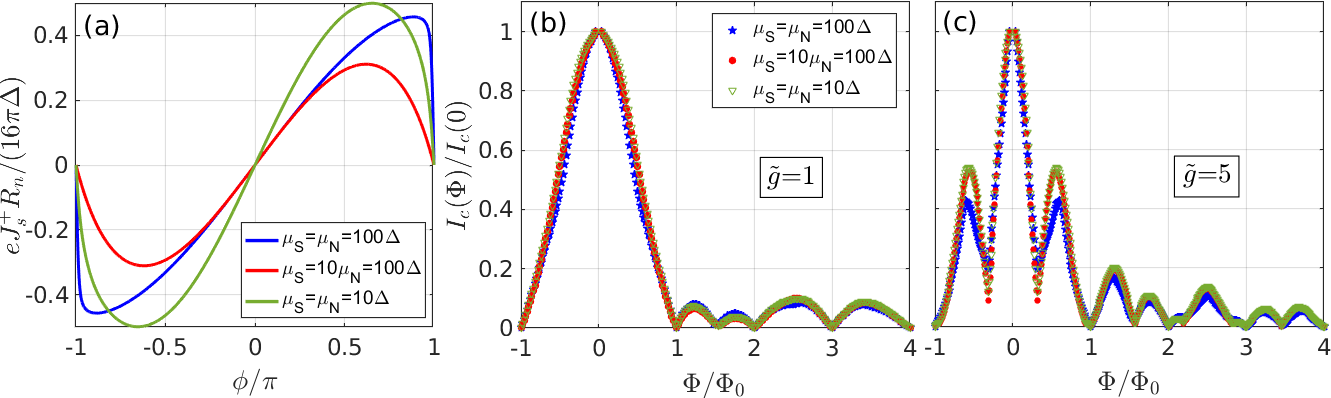}

\caption{Current-phase relations for positive chirality in the absence of magnetic
fields (a) and Fraunhofer patterns for $\tilde{g}=1$ (b) and $5$
(c) in junctions with chemical potentials $\mu_{S}=\mu_{N}=100\Delta,$
$\mu_{S}=10\mu_{N}=100\Delta$, and $\mu_{S}=\mu_{N}=10\Delta$, respectively.
Other parameters are $L=1/(100\Delta)$ and $k_{B}T=0.01\Delta$. }

\label{Fig:Fraunhofer-pattern-mu}
\end{figure}

\begin{figure}[H]
\centering

\includegraphics[width=16cm]{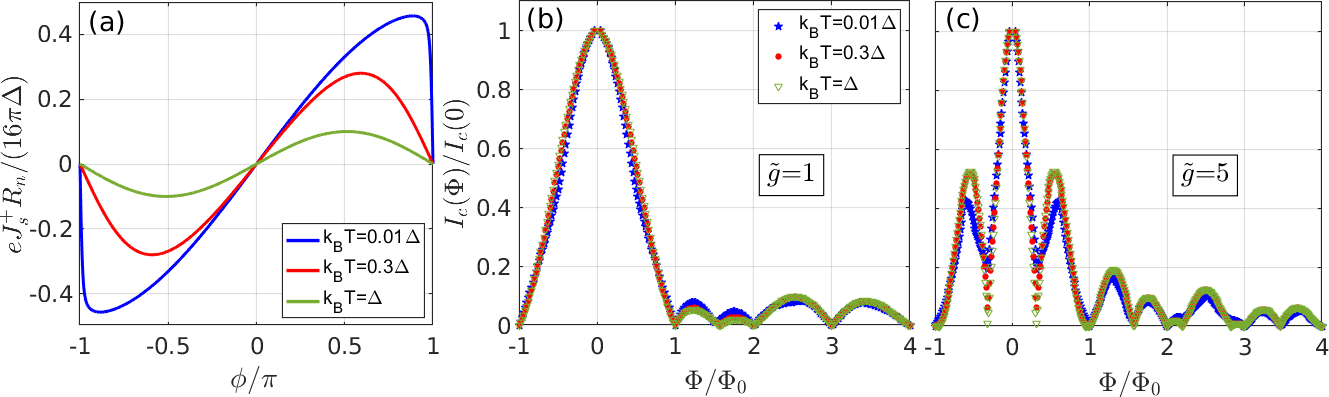}

\caption{Current-phase relations for positive chirality in the absence of magnetic
fields (a) and Fraunhofer patterns for $\tilde{g}=1$ (b) and $5$
(c) at temperature $k_{B}T=0.01\Delta,\ 0.3\Delta$, and $\Delta$,
respectively. Other parameters are $\mu_{S}=\mu_{N}=100\Delta$ and
$L=1/(100\Delta)$. }

\label{Fig:Fraunhofer-pattern-T}
\end{figure}

As shown in Figs.\ \ref{Fig:Fraunhofer-pattern-1}, \ref{Fig:Fraunhofer-pattern-mu}
and \ref{Fig:Fraunhofer-pattern-T}, although the CPR is sensitive
to junction lengths, low temperatures, and chemical potentials, the
corresponding Fraunhofer patterns are almost the same. For $\tilde{g}\neq0$,
the supercurrents of opposite chirality vary differently as increasing
the magnetic field. As a consequence, an anomalous Fraunhofer pattern
occurs as the interference of the two different supercurrents. Therefore,
the anomalous Fraunhofer pattern is an indication of the finite chirality
Josephson currents in the system. At large temperatures, the sinusoidal
form of CPR can be observed. Thus, all the minima drop to zero {[}Fig.\ \ref{Fig:Fraunhofer-pattern-T}(b,
c){]}.

Since different choices of $J_{s}^{\text{tot}}(\phi)$ yield more
or less the same Fraunhofer pattern, we can assume the usual sinusoidal
CPR for each chirality and obtain a simple formula
\begin{align}
\dfrac{I_{c}(\Phi)}{I_{c}(0)} & =\dfrac{\Phi_{0}}{\pi\Phi}\left|\sin\Big(\dfrac{\pi\Phi}{\Phi_{0}}\Big)\cos\Big(\dfrac{\tilde{g}\Phi}{\Phi_{0}}\Big)\right|.\label{eq:fitting-expression}
\end{align}
Although the true CPR is usually skewed forward, Eq.\ (\ref{eq:fitting-expression})
captures the correct Fraunhofer patterns quite well (Fig. 4 in the Letter).
From this analogy, the two frequencies of $\Phi/\Phi_{0}$ can be extracted as $2/(1\pm\tilde{g}/\pi)$, respectively. $I_{c}(\Phi)$ vanishes not only at every nonzero integer $\Phi/\Phi_{0}\text{\ensuremath{\in}}\{\pm1,\pm2,...\}$,
but also at $\Phi/\Phi_{0}=(n+1/2)\pi/\tilde{g}$, $n\in\mathbb{Z}$. The skewed form in $J_{s}^{\pm}(\phi)$, however, removes the zeros of $I_{c}(\Phi)$ at $\Phi/\Phi_{0}=(n+1/2)\pi/\tilde{g}$ and leaves finite minima there.

\end{document}